\documentstyle[12pt]{article}

\advance\textheight by \topskip
\begin{document}
\tolerance=100000
\input feynman
\thispagestyle{empty}
\setcounter{page}{0}

\newcommand{\be}{\begin{equation}}
\newcommand{\ee}{\end{equation}}
\newcommand{\br}{\begin{eqnarray}}
\newcommand{\er}{\end{eqnarray}}
\newcommand{\ba}{\begin{array}}
\newcommand{\ea}{\end{array}}
\newcommand{\bi}{\begin{itemize}}
\newcommand{\ei}{\end{itemize}}
\newcommand{\bn}{\begin{enumerate}}
\newcommand{\en}{\end{enumerate}}
\newcommand{\bc}{\begin{center}}
\newcommand{\ec}{\end{center}}
\newcommand{\ul}{\underline}
\newcommand{\ol}{\overline}
\newcommand{\eebbww}{$e^+e^-\rightarrow b\bar b W^+W^-$}
\newcommand{\bb}{$ b\bar b \ $}
\newcommand{\ttb}{$ t\bar t \ $}
\newcommand{\ar}{\rightarrow}
\newcommand{\sm}{${\cal {SM}}\ $}
\newcommand{\Dir}{\kern -6.4pt\Big{/}}
\newcommand{\Dirin}{\kern -10.4pt\Big{/}\kern 4.4pt}
\newcommand{\DDir}{\kern -7.6pt\Big{/}}
\newcommand{\DGir}{\kern -6.0pt\Big{/}}

\def\Ord{\buildrel{\scriptscriptstyle <}\over{\scriptscriptstyle\sim}}
\def\OOrd{\buildrel{\scriptscriptstyle >}\over{\scriptscriptstyle\sim}}
\def\pl #1 #2 #3 {{\it Phys.~Lett.} {\bf#1} (#2) #3}
\def\np #1 #2 #3 {{\it Nucl.~Phys.} {\bf#1} (#2) #3}
\def\zp #1 #2 #3 {{\it Z.~Phys.} {\bf#1} (#2) #3}
\def\pr #1 #2 #3 {{\it Phys.~Rev.} {\bf#1} (#2) #3}
\def\prep #1 #2 #3 {{\it Phys.~Rep.} {\bf#1} (#2) #3}
\def\prl #1 #2 #3 {{\it Phys.~Rev.~Lett.} {\bf#1} (#2) #3}
\def\mpl #1 #2 #3 {{\it Mod.~Phys.~Lett.} {\bf#1} (#2) #3}
\def\rmp #1 #2 #3 {{\it Rev. Mod. Phys.} {\bf#1} (#2) #3}
\def\xx #1 #2 #3 {{\bf#1}, (#2) #3}
\def\preprint{{\it preprint}}

\begin{flushright}
{\large DFTT 23/94}\\
{\large DTP/94/42}\\
{\rm June 1994\hspace*{.5 truecm}}\\
\end{flushright}

\vspace*{\fill}

\begin{center}
{\Large \bf Heavy
Higgs production and decay via $e^+e^-\rightarrow Z^0 H^0
\rightarrow b\bar bW^+W^-$
and irreducible backgrounds at the Next Linear
Collider.\footnote{Work supported in part by Ministero
dell' Universit\`a e della Ricerca Scientifica.\\[4. mm]
E-mails: Ballestrero,Maina,Moretti@to.infn.it;
Stefano.Moretti@durham.ac.uk.}}\\[2.cm]
{\large Alessandro Ballestrero$^a$, Ezio Maina$^a$,
Stefano Moretti$^{a,b}$}\\[0.5 cm]
{\it a) Dipartimento di Fisica Teorica, Universit\`a di Torino,}\\
{\it and I.N.F.N., Sezione di Torino,}\\
{\it Via Pietro Giuria 1, 10125 Torino, Italy.}\\[0.5cm]
{\it b) Department of Physics, University of Durham,}\\
{\it South Road, Durham DH1 3LE, United Kingdom.}\\[0.75cm]
\end{center}

\vspace*{\fill}

\begin{abstract}
{\normalsize
\noindent
The complete matrix element for $e^+e^-\ar b\bar bW^+W^-$ has been
computed at tree--level and applied to
$Z^0H^0$ production followed by $Z^0\ar b\bar b$ and $H^0\ar W^+W^-$,
keeping into account all irreducible
backgrounds, which are dominated by \ttb production,
at the Next Linear Colliders. We find that, depending
on the center of mass energies and
on the search strategies, this channel can be useful
for the study of the
parameters of the Standard
Model Higgs boson over the most part of the heavy mass range.}
\end{abstract}

\vspace*{\fill}
\newpage

\subsection*{Introduction}

The Standard Model (\sm \hspace{-.4em})
postulates the existence of a massive scalar particle,
the Higgs boson $H^0$, whose role is crucial in generating
the spontaneous symmetry breaking of the $SU(2)_L\times U(1)_Y$
gauge group of the electroweak interactions, and in ensuring
the renormalizability of the theory.
Except for an upper bound of approximately 1 TeV,
which can be derived from perturbative
unitarity arguments \cite{unitarity}, the model does not
make any prediction on the mass $M_{H^0}$ of such a particle.\par
At present, a lower limit can be extracted
from LEP I ($\sqrt s_{ee}=M_{Z^0}$) experiments:
from the results of searches for $e^+e^-\ar
Z^0\ar Z^{0*}H^0$ events, one derives the bound
$M_{H^0}\OOrd 60~{\rm {GeV}}$ \cite{limSM}.\par
Extensive studies have been carried out on the feasibility of
discovering
the Higgs particle by the next generation of high energy
machines, both at  $pp$ \cite{guide,LHC,SSC} and at
$e^+e^-$ colliders \cite{guide,LepII,NLC,ee500,LC92,JLC}.\par
The mass region $M_{H^0} < 80\div 90$ GeV can be studied at LEP II
($\sqrt s_{ee}=160\div200$ GeV) whereas
a Higgs with a larger mass will be searched for
at $pp$ colliders like LHC ($\sqrt s_{pp}=14$ TeV)
or at $e^+e^-$ accelerators like NLC ($\sqrt s_{ee}=300\div 1000$ GeV).
\par
At LHC the mass range 80 GeV $\Ord M_{H^0}
\Ord 130$ GeV results the most difficult to study since
in this case the Higgs boson mainly
decays to $b\bar b$ pairs and the QCD background is huge.
However, recent studies have shown that
it is possible to detect the $H^0$
in the $\gamma\gamma$ decay mode \cite{gamgam},
via the associated production with a $W^\pm$ boson \cite{gny,wh}
or a $t\bar t$ pair \cite{rwnz,tth}. For $M_{H^0} \OOrd 130$ GeV,
the Higgs can be discovered in the ``gold-plated'' four--lepton mode
$H^0\rightarrow Z^0Z^0\rightarrow \ell\bar\ell\ell\bar\ell$
\cite{LHC,SSC}.\par
At NLC, with $\sqrt s_{ee}=300\div 500$ GeV,
the Higgs detection can be achieved
over the whole intermediate mass range
$M_{Z^0}\Ord M_{H^0}\Ord 2 M_{W^\pm}$ \cite{BCDKZ}. The two main
production mechanisms are the Bjorken bremsstrahlung reaction  $e^+e^-
\rightarrow Z^{0*}\rightarrow Z^0H^0$ \cite{bremSM} which dominates
below $\sqrt s_{ee}=500$
and the fusion processes
$e^+e^-\rightarrow \bar\nu_e\nu_eW^{\pm*}
W^{\mp*}(e^+e^-Z^{0*}Z^{0*})\rightarrow\bar\nu_e\nu_e
(e^+e^-)H^0$ \cite{fusionSM} which dominate at larger energies.
At $\sqrt s_{ee}\OOrd500$ GeV a heavy Higgs, in addition than in
the $4\ell$--mode, can be detected
in the four--jet modes
$H^0\rightarrow W^\pm W^\mp,Z^0Z^0\rightarrow jjjj$
\cite{BCKP,4jet}.\par
Recently, the $b$--tagging capabilities achieved by vertex
detectors have suggested
the possibility of Higgs searches through new signatures.
For example,
it has been shown \cite{btagg} that with the $b$--tagging
performance
\cite{SDC} foreseen for LHC experiments it may be possible to see
the $H^0$ in the $t\bar tH^0$ production channel, with one $t$
decaying semileptonically, followed by
$H^0\rightarrow b\bar b$, for 80 GeV $\Ord
M_{H^0} \Ord$ 130 GeV,
provided that $m_t\OOrd 130$ GeV.\par
It is generally expected, even in these financially
troubled times for particle physics,
that LHC will begin to operate approximately ten years from now, while
the projects for a NLC--type of $e^+e^-$ machine are still at a very
preliminary stage. Therefore it is reasonable to assume that the Higgs,
if it exists at all, will be discovered, and that its mass
will be first measured, at a $p p$ collider.
However it will be impossible to study in detail all couplings
between the Higgs boson and  the other particles of the \sm at a
hadron machine.
For this purpose, as well as for a full analysis and a precise determination
of the parameters of the $top$
quark, a high--energy linear $e^+e^-$ machine will be essential.\par
At NLC the Higgs mass can be extracted from missing mass analyses or from a
direct measurement of the decay products \cite{ee500}.
The first method requires a measurement of the momentum of the $Z^0$.
This can be done easily for the $e^+e^-$ and $\mu^+\mu^-$ decays
but, with the design
total luminosity of $10$ $fb^{-1}$,
at the price of very small statistics \cite{BCDKZ}.
Alternatively one can exploit
the much larger $b\bar b$ decay mode and the capabilities of vertex detectors.
We emphasize that the channel $Z^0\ar b\bar b$
might well be one of best ways to detect the $Z^0$.
In fact this mode is free from backgrounds coming
from $W^\pm$ decays, and
has a branching ratio which is
about five times larger than the corresponding ones
to $Z^0\ar e^+e^-$ and $Z^0\ar\mu^+\mu^-$,
comparable to the fraction
of invisible decays $Z^0\rightarrow \nu\bar\nu$
(the possibility of studying Higgs production using these latter
has been examined in ref. \cite{BCKP}). \par
The missing mass method has the unique feature of being
independent of assumptions on the $H^0$ decay modes. The
$H^0\rightarrow W^\pm W^\mp$ decays can be easily sorted out no matter
how the $W$'s decay and the corresponding branching ratio
can be directly measured.
In order to obtain a precise measurement of the Higgs
mass the beam energy spread resulting from
bremsstrahlung and beamsstrahlung has to be taken into account.
The missing mass distribution will peak at a larger value
than the true Higgs mass, but this can be easily corrected for
once the parameters of the machine are known.
For narrow--band beam designs as DLC and TESLA \cite{barklow}
beamsstrahlung effects are much smaller than those
of initial--state--radiation.
Photon emission lowers the center--of--mass energy
at which the collisions between electrons and positrons
take place,
decreasing the cross sections at thresholds and,
on the contrary, increasing production rates
at higher energies since the bulk of events are produced
through $s$--channel annihilation and therefore scale as $s^{-1}$.
For the mentioned designs the global effect is at most about
20\%. Since we neglect bremsstrahlung and beamsstrahlung
both the number of events and their statistical significance
will be slightly higher than we predict.\par
The direct reconstruction of the Higgs decay products requires both
$W$'s to decay hadronically, which halves the statistics. It is however
the only available mean if the $Z$ decays to $\nu\bar\nu$ or for
Higgs produced in fusion processes. The mass determination will suffer
from the uncertainties in the measurement of jet energies but this
can be somewhat
improved exploiting the fact that the invariant mass of two pairs of jets
must reconstruct the $W$ mass.\par
In principle one could also consider the six--jet channel,
$e^+e^-\ar Z^0H^0 \ar Z^0 W^+W^- \ar 6j$, looking for a peak in the
invariant mass distribution of pairs of vector bosons. This method
however suffer from a large combinatorial background.
For each event three different pairs can be formed since it is unlikely
that the $Z$ can be distinguished from the $W$'s, solely on the basis
of their mass, when all three
decay hadronically, the mass difference being comparable to the
expected resolution.\par
Using the full matrix element for the process
$e^+e^-\ar b\bar bW^+W^-$ we are able
to study the production of a heavy \sm\ Higgs (i.e., $M_{H^0}\ge 2M_{W^\pm}$)
through the Bjorken bremsstrahlung reaction
$e^+e^-\ar Z^0H^0$, followed by the
decays $Z^0\ar b\bar b$ and $H^0\ar W^+W^-$, and of all irreducible
backgrounds.
The main background is
due to $t\bar t\ar b\bar b W^+W^-$ production and decay
which has a much larger cross section than $e^+e^-\ar Z^0H^0$
\cite{sigmatop,sigmaH}.
Therefore it is essential in order to assess the observability of the latter
to have a complete description of the former and of the interference between
the two.\par
The plan of the paper is as follows. In section II we
give details on the calculation, while in section III we present
and discuss the results. Finally, section IV is devoted to the conclusions.

\subsection*{Calculation}

We have computed the matrix element of the
process $e^+e^-\ar b \bar b W^+W^-$ using
the method of ref. \cite{method}.
As a check, we have performed the calculation
also by means of the formalism \cite{hz},
and compared the corresponding
{\tt FORTRAN} codes.
The $b(\bar b)$ quark has been treated as a massive stable particle, while
the widths of all virtual $t(\bar t)$ quark, $W^\pm$, $Z^0$ and
$H^0$ bosons
have been included in our calculations.
For the top and the Higgs width we have adopted
their tree--level expressions,
while
we have not included the effects of the width of the final state $W$'s
\cite{W_width}.
For further
details on the calculations, we refer to ref.~\cite{noieebbww}.\par
The matrix element squared has to be integrated with some care
in order to control the interplay among the various peaks
which appear in correspondence to the possible resonances.
We have divided the full set of Feynman diagrams, which are shown
in fig.1, in eight
subset, according to the different resonant structures,
as follows:
\br
t\bar t\ar (bW^+)(\bar bW^-) & &{\rm diagram~\#~2}, \\
t\ar bW^+ & &{\rm diagrams~\#~1,7,17},                \\
\bar t\ar \bar b W^- & &{\rm diagrams~\#~3,8,16},       \\
Z^0H^0\ar (b\bar b)(W^+W^-) & &{\rm diagram~\#~26},     \\
Z^0\ar b\bar b & &{\rm diagrams~\#~4,5,6,11\div 15,18,19}, \\
H^0\ar W^+W^- & &{\rm diagrams~\#~20,21},               \\
H^0\ar b\bar b & &{\rm diagrams~\#~22\div 25,27}.          \er
Diagrams \# 9,10 and diagrams 4,6,11$\div$15,18,19 with an
intermediate photon constitute the eighth (non--resonant) channel.
If we indicate by $M_i$ the sum of the diagrams
in the $i$--th channel, one has
\be\label{sum} M_{tot}=\sum_{i=1}^{8} M_i,\ee
where $M_{tot}$ is the total Feynman amplitude.
In squaring equation (\ref{sum}) we take the combinations
\be\label{m1} {\cal T}_{1}=|M_1|^2,                       \ee
\be\label{m2} {\cal T}_{2}=|M_2|^2+2Re[M_1M_2^*], \ee
\be\label{m3} {\cal T}_{3}=|M_3|^2+2Re[M_1M_3^*], \ee
\be\label{m4} {\cal T}_{4}=|M_4|^2,                       \ee
\be\label{m5} {\cal T}_{5}=|M_5|^2+2Re[M_4M_5^*], \ee
\be\label{m6} {\cal T}_{6}=|M_6|^2+2Re[M_4M_6^*], \ee
\be\label{m7} {\cal T}_{7}=|M_7|^2,                       \ee
\be\label{m8} {\cal T}_{8}=|M_8|^2
               + \mbox{all remaining interference terms}.    \ee
Obviously
\be |M_{tot}|^2=\sum_{i=1}^{8} {\cal T}_i.\ee
\par
In the end, we have separately integrated
the various contributions (\ref{m1})--(\ref{m8})
with VEGAS \cite{vegas}, using
$y_X = \tan^{-1}\left( \frac{Q^2-M^2_X}{M_X\Gamma_X}\right)$
as integration variable when the $X$--resonance was
present.\par
Finally, concerning the numerical part of our work,
we have used the following values:
$M_{Z^0}=91.1$ GeV, $\Gamma_{Z^0}=2.5$ GeV,
$\sin^2 (\theta_W)=0.23$, $M_{W^\pm}=M_{Z^0}\cos (\theta_W)=79.9$ GeV,
$\Gamma_{W^\pm}=2.2$ GeV,
$m_b=5.0$ GeV and $\alpha_{em}= 1/128$.\par

\subsection*{Results}
If one looks at the $b  \bar b W^+W^-$ final state in the
$M_{b\bar b}$ versus $M_{W^+W^-}$ plane the events
from the bremsstrahlung reaction will concentrate in a single
blot whose size is determined by the experimental
reconstruction incertainties. On the contrary one expects
the background events, mostly $t\bar t$ production,
to fill a large region. It is
however possible, for a given
$top$ mass, $m_t$, to find
values of the Higgs mass for which the two
double--differential distributions
do no overlap. In these cases \ttb production
can be very simply distinguished from
Higgs production and detecting the $H^0$ is only a matter
of event rate. In fig.2
the boundaries of the $d\sigma/dM_{b\bar b}/dM_{W^+W^-}$
distribution in the plane
$(M_{b\bar b},M_{W^+W^-})$ are presented
for $e^+e^-\rightarrow t\bar t$ events
for several values of $m_t$ at $\sqrt s=350$ GeV
and $\sqrt s=500$ GeV.
It can be seen that, for $M_{b\bar b}=M_{Z^0}$, there is indeed
a range of Higgs masses, between threshold
and a maximum value which decreases with increasing $m_t$, for which
Higgs production is effectively free of any background from
$t\bar t$ production.
For the special case $\sqrt s=350$ GeV and $m_t= 170$ GeV there is also
a small region at the high end of the range of Higgs masses which produce
an appreciable number of events which is free from \ttb background.\par
In fig.3 and fig.4 we present the invariant mass distribution of the
system recoiling against the \bb pair at $\sqrt s=350$ GeV
and $\sqrt s=500$ GeV, respectively, for $m_t= 150$ GeV and $m_t= 175$ GeV.
We have required $|\cos\theta_{b\bar b}|<0.8$, in order to have events
which can be tagged by the microvertex detectors,
and $|M_{Z^0}-M_{b\bar b}|<10$ GeV. These figures have been obtained
from the full matrix element. A separation of signal from
background is strictly speaking impossible, however the broad structure
on which the Higgs peaks rest is largely independent of $M_H$
and one can identify it with the irreducible background.
This distribution is almost flat within the range where \ttb production
is kinematically allowed, apart from a narrow region
at the boundaries where it rapidly decreases to
essentially zero.\par
The precise composition of the total
background depends on the strategy
adopted for $b$--tagging. If only one of the two $b$--jets
is required to be identified
there is, in addition to the irreducible background,
a reducible contribution arising from the possible
misidentification of a $c$--quark as a $b$--quark.
If $\epsilon_{b(c)}$ is the probability for a $b(c)$ quark
to satisfy a given set of tagging requirements,
the probability of
tagging at least one $b(c)$ of a $b\bar b (c \bar c )$ pair is
$P_{b(c)} = (1-(1-\epsilon_{b(c)})^2)$.
The fraction of $W$ decays involving a $c$--quark is
large $B_{Wc}\equiv B(W^+\ar c\bar s)\approx
B(W^+\ar u\bar d)\approx 33\%$.
Therefore, the probability that a $W^+W^-$ pair results in at least
one tag
is $P_{WW} = B_{Wc}^2 (4\epsilon_c -\epsilon_c^2)$. Taking
for example
$\epsilon_{b}=33\%$ and $\epsilon_{c}=5\%$ \cite{tag}
and defining $\sigma_0 = \sigma (e^+e^-\ar ZH) \times B(H\ar WW)$
the reducible background from the reaction
$e^+e^-\ar ZH$ is approximately equal to
$\sigma_0 \times (B(Z\ar udsc) \times P_{WW} + B(Z\ar c)\times P_c )\approx
\sigma_0 \times 2.5\%$.
An additional source of
fake tags are the $WW$ pairs produced in association with
a \bb pair. We can estimate the corresponding cross section
multiplying by $P_{WW}\approx 2\%$ the integral of the distributions in
fig.3 and 4 in a 10 GeV bin centered around the Higgs nominal mass.
The largest contribution is only about 0.5 $fb$
for $\sqrt s=350$ GeV, $m_t= 150$ GeV and $M_H= 185$ GeV.
The sum of reducible and irreducible background
is to be compared with the signal which is
$\sigma_0 \times B(Z\ar b) \times P_b \approx
\sigma_0 \times 8.\%$.
An alternative option, which has the advantage of drastically
reducing the irreducible
background, is to impose rather loose
identification requirements on both $b$--jets,
exploiting the fact that it is less likely
to mistake a $cs$ pair for a $bb$ pair
than it is to mistake a single $c$--quark for a $b$--quark.
In ref.\cite{tag} it has been shown that with vertex detectors comparable
with those already in operation it is possible to achieve an efficiency for
detecting \bb pairs equal to .466 with an efficiency for $cs$ pairs of
.025.\par
Neglecting the complications due to the irreducible
background for the time being,
a rough estimate of the statistical significance of the signal is
given by the ratio:
\be\label{ratio}
\frac {\rm S}{\sqrt {\rm B}} = \sqrt{P_b {\cal L} } \;
\frac {\sigma (e^+e^-\ar ZH\ar b\bar b W^+W^-)}
{\sqrt{ \sigma (e^+e^-\ar b\bar b W^+W^-) -
\sigma (e^+e^-\ar ZH\ar b\bar b W^+W^-)}}
\ee
where $\cal L$ is the luminosity and we integrate all cross sections,
with the cuts used to produce fig.3 and 4,
in a 10 GeV bin centered around the Higgs nominal mass.
The ratio (\ref{ratio}), together with the expected number of events,
is presented in table I,
for ${\cal L} = 10 \; fb^{-1}$ and
$\epsilon_b=1/3$. In most cases,
when the \ttb background is substantial,
the significance is quite small and insufficient
to detect the Higgs signal.\par
It is obvious that many features can be helpful in
distinguishing signal from background events.
Neglecting brems\-strah\-lung and
beams\-strah\-lung effects in \ttb pro\-duction,
the energy of the two heavy quarks equals the beam energy.
Moreover for each $W$ one can construct two invariant masses
coupling the $W$ in turn to both $b$'s and one of these quantities
must be equal to $m_t$. Since the number of signal
events is not too small, it is reasonable to try to increase the signal
to background ratio with a more stringent set of cuts.
In order to reduce as little as possible the signal we have
considered events for which
$|M_{Z^0}-M_{b\bar b}|<10$ GeV and  have required
that one of the $W$'s, say the $W^+$, failed to reproduce the kinematics
of a \ttb
final state when coupled with either of the two $b$'s, namely that
$m_t-10~{\rm{GeV}}>|M_{W^+b(W^+\bar b)}|>m_t+10~{\rm{GeV}}$
and $E_{beam}-10~{\rm{GeV}}>|E_{W^+}+E_{b(\bar b)}|>
E_{beam}+10~{\rm{GeV}}$.
The resulting cross sections,
integrated over the window $|M_{H^0}-M_{W^+W^-}|<10$ GeV, are presented
in table II and III. The background has been reduced to a very small level
while only between 10\% and 30\% of the signal has been lost and
a reasonable number of events are expected over the whole
range of Higgs masses we have examined.
The proposed selection criteria are particularly
convenient, since only one of the two
$W$'s is required to decay hadronically, therefore about 85\%
of the $WW$ decays of the Higgs are retained. This reduction factor
and the $b$--tagging efficiency are not included
in the cross section presented in table II and III.\par

\subsection*{Conclusions}
We have studied, using full matrix element for the process
$e^+e^-\ar b\bar bW^+W^-$,
the production of a \sm\ Higgs with $M_{H^0}\ge 2M_{W^\pm}$
through the bremsstrahlung reaction $e^+e^-\ar Z^0H^0$, followed by the
decays $Z^0\ar b\bar b$ and $H^0\ar W^+W^-$, and the
corresponding irreducible
background which is dominated by \ttb production.
Selecting events containing a \bb pair compatible with a
decay of $Z^0$, it has been shown
that there are values of the Higgs and the $top$ mass for which a simple
cut on the invariant
mass of the system recoiling against the \bb pair is sufficient to
completely eliminate the irreducible background. When the distributions
$d\sigma/dM_{b\bar b}/dM_{W^+W^-}$ in the plane
$(M_{b\bar b},M_{W^+W^-})$ of signal and background events overlap,
the statistical significance $S/\sqrt{B}$ of the signal is
not sufficient, in general, to unambigously establish the presence
of the Higgs with a simple missing mass analysis. Further cuts,
based on the kinematics of \ttb production,
can however drastically reduce the background while mantaining
an acceptable number of events from Higgs production.
Therefore we conclude that the production of a heavy \sm\ Higgs
in association with a $Z^0$
can be observed at NLC in the decay channels $Z^0\ar b\bar b$
and $H^0\ar W^+W^-$.

\vfill
\newpage

\thispagestyle{empty}

\subsection*{Table Captions}

\begin{description}

\item[tab.I  ] Expected number of signal
and background events and their statistical significance
at $\sqrt s=350$ GeV and $\sqrt s=500$ GeV
for a selection of Higgs masses after
the cuts:
$|M_{Z^0}-M_{b\bar b}|<10$ GeV and $|\cos\theta_{b\bar b}|<0.8$.
We assume that only one $b$-jet is tagged with efficiency
$\epsilon_b=1/3$. The luminosity is taken to be
${\cal L} = 10~fb^{-1}$.

\item[tab.II  ] Cross sections for
$e^+e^-\rightarrow b\bar b W^+W^-$
at $\sqrt s=350$ GeV, for five
different values of $M_{H^0}$.
The first column is the resonant contribution
$e^+e^-\rightarrow Z^0H^0\rightarrow b\bar b W^+W^-$
in the absence of cuts.
The second and third column give the resonant cross section
and the cross section obtained from all
diagrams without Higgs, assuming  $m_t=150$ GeV,
after the following set of cuts:
$|M_{Z^0}-M_{b\bar b}|<10$ GeV, $|M_{H^0}-M_{W^+W^-}|<10$ GeV,
$m_t-10~{\rm{GeV}}>|M_{W^+b(W^+\bar b)}|>m_t+10~{\rm{GeV}}$
and $E_{beam}-10~{\rm{GeV}}>|E_{W^+}+E_{b(\bar b)}|>
E_{beam}+10~{\rm{GeV}}$.

\item[tab.III  ] Cross sections for
$e^+e^-\rightarrow b\bar b W^+W^-$
at $\sqrt s=500$ GeV, for five
different values of $M_{H^0}$.
The first column is the resonant contribution
$e^+e^-\rightarrow Z^0H^0\rightarrow b\bar b W^+W^-$
in the absence of cuts.
The second and third column give the resonant cross section
and the cross section obtained from all
diagrams without Higgs, assuming  $m_t=150$ GeV,
after the following set of cuts:
$|M_{Z^0}-M_{b\bar b}|<10$ GeV, $|M_{H^0}-M_{W^+W^-}|<10$ GeV,
$m_t-10~{\rm{GeV}}>|M_{W^+b(W^+\bar b)}|>m_t+10~{\rm{GeV}}$
and $E_{beam}-10~{\rm{GeV}}>|E_{W^+}+E_{b(\bar b)}|>
E_{beam}+10~{\rm{GeV}}$.

\end{description}

\vfill
\newpage
\thispagestyle{empty}

\subsection*{Figure Captions}

\begin{description}

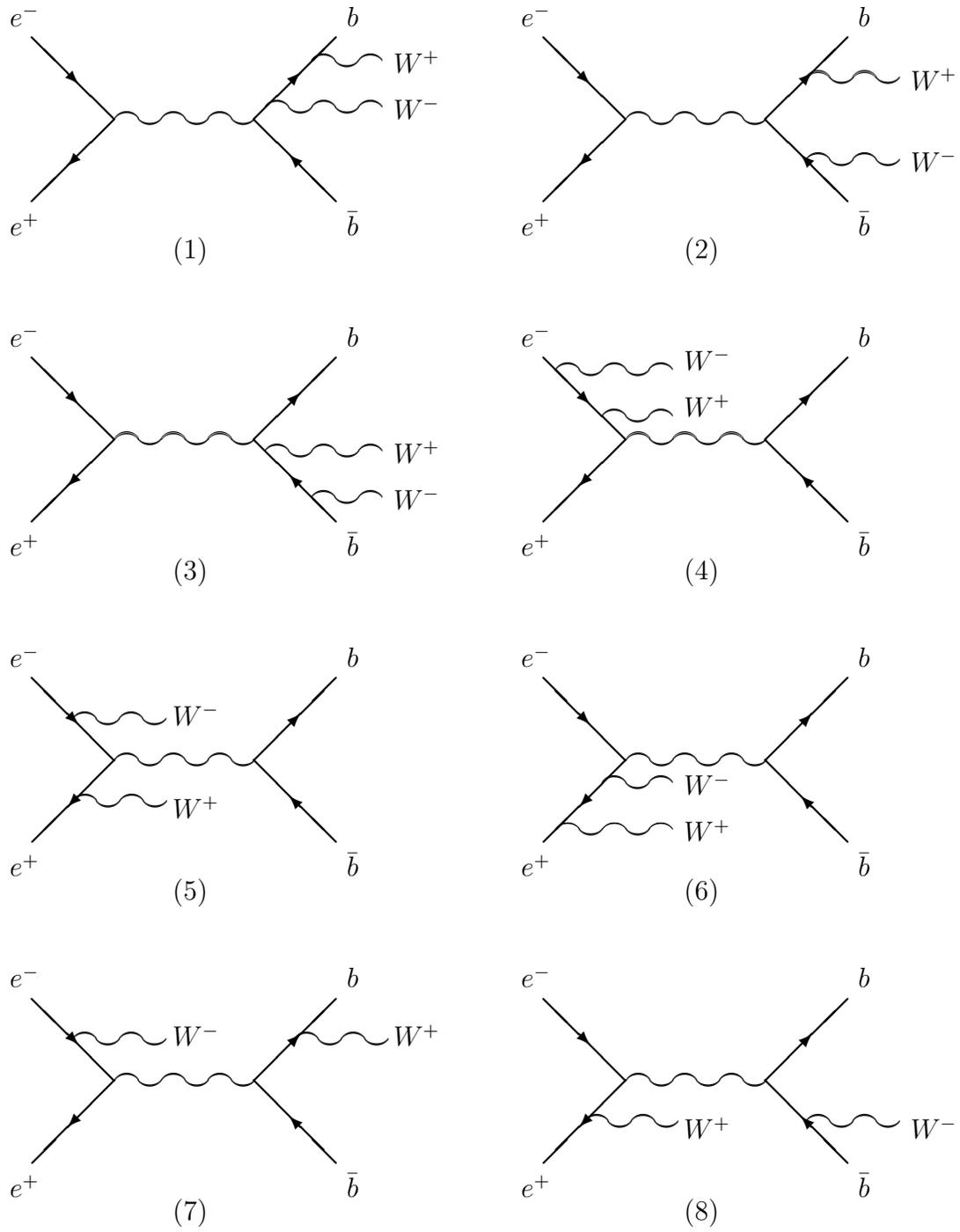
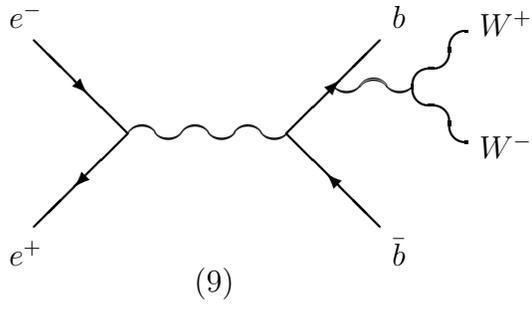
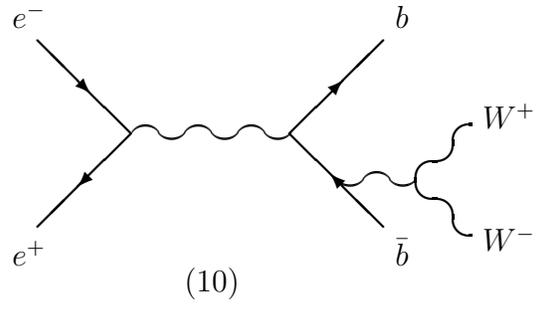
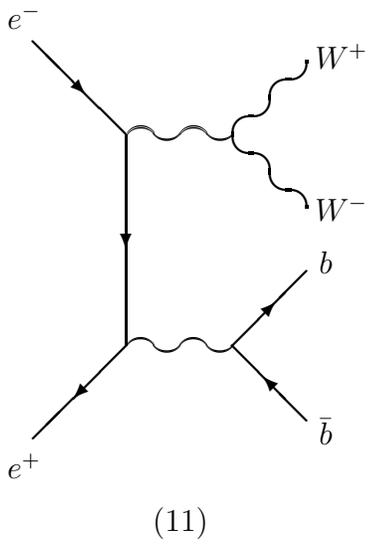
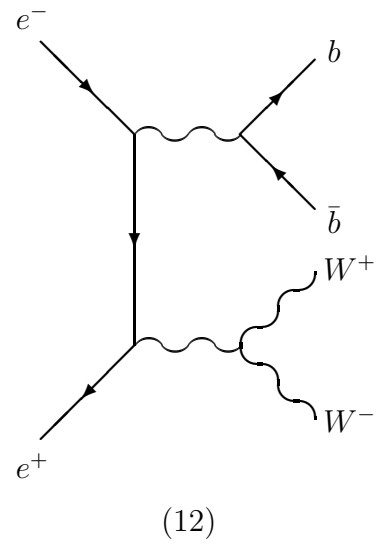
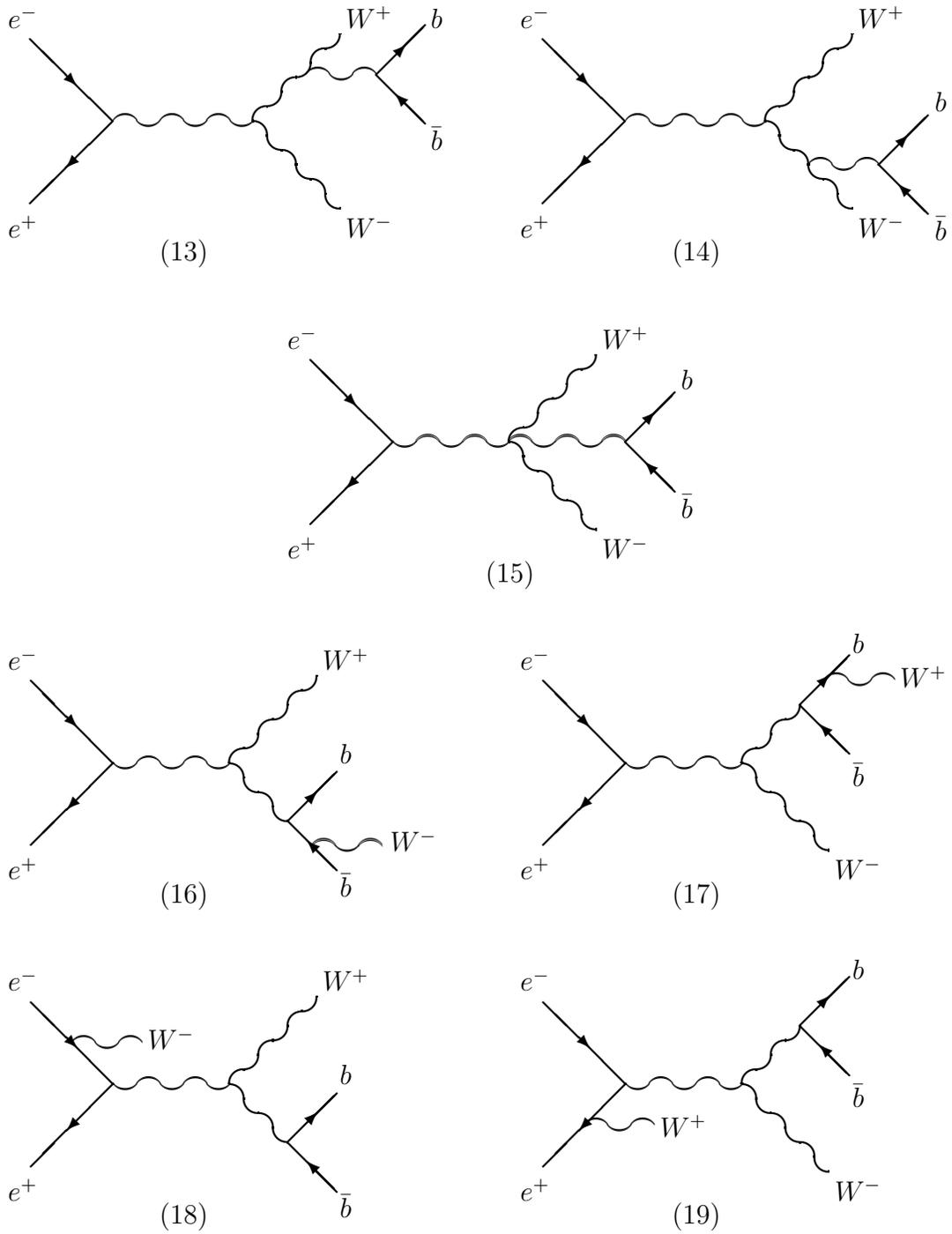
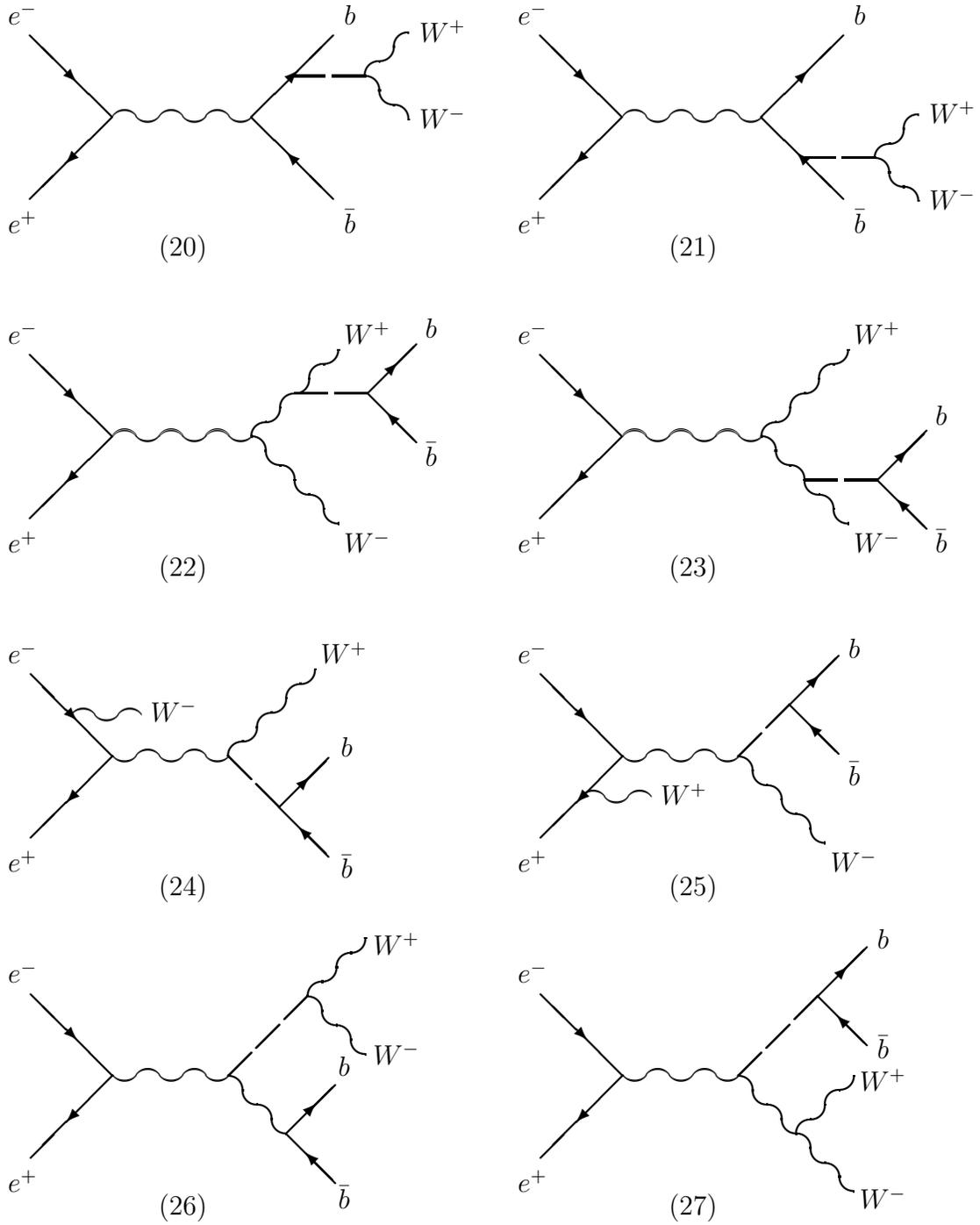
\item[fig.1 ] Feynman diagrams contributing in the lowest order to
$e^+e^-\rightarrow b\bar b W^+W^-$.
Internal wavy lines represent a $\gamma$, a $Z^0$ or a $W^\pm$,
as appropriate. Internal dashed lines represent a Higgs boson.

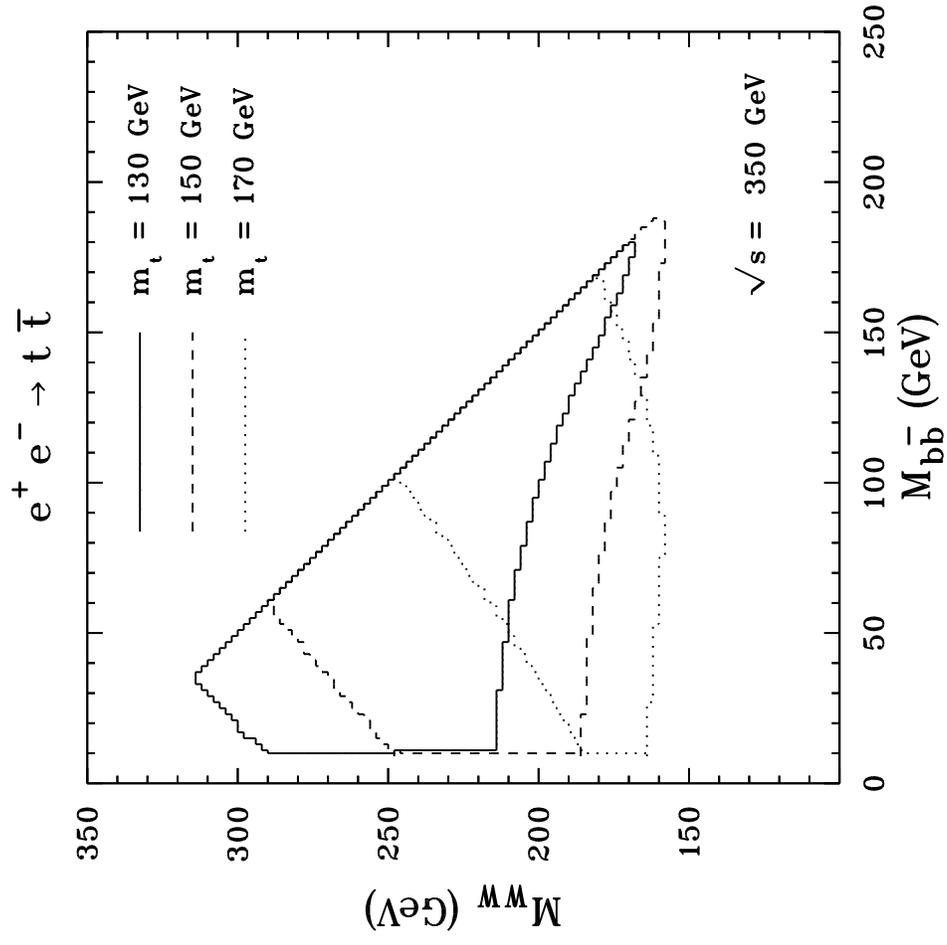
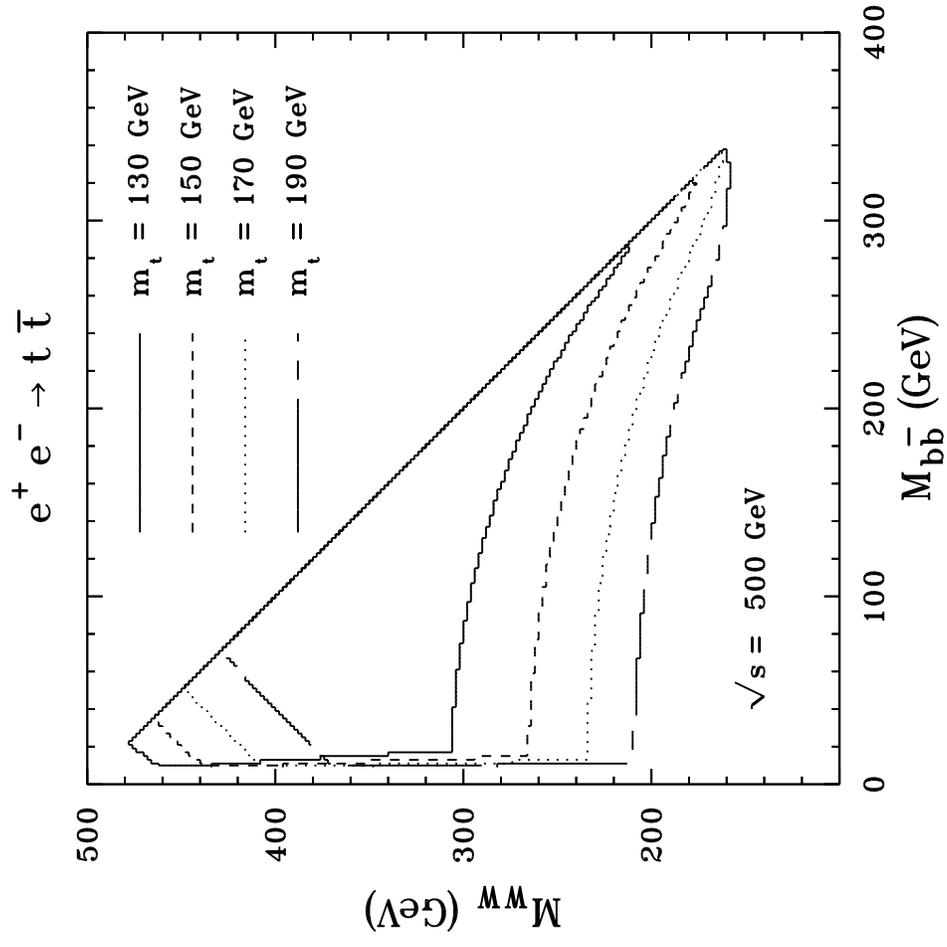
\item[fig.2 ] The boundaries of the double differential distribution
$d\sigma/dM_{b\bar b}/dM_{W^+W^-}$ in the plane
$(M_{b\bar b},M_{W^+W^-})$ for $e^+e^-\rightarrow t\bar t$ events
in the narrow width approximation,
at $\sqrt s=350$ GeV and $\sqrt s=500$ GeV,
for different values of $m_t$, without cuts.

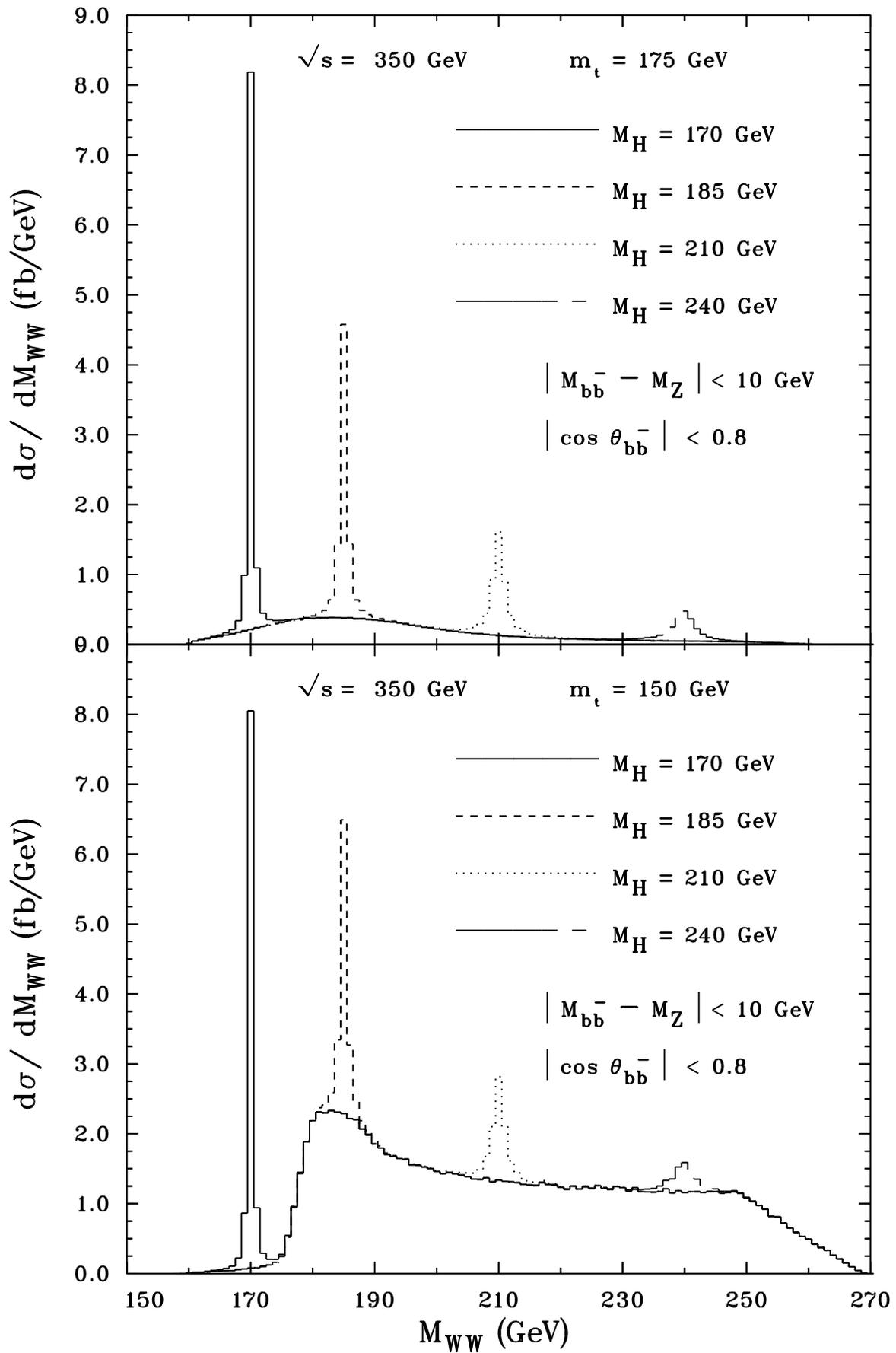
\item[fig.3] The differential distribution
$d\sigma/dM_{W^+W^-}$ for
$e^+e^-\rightarrow b\bar b W^+W^-$ (full matrix element
with all Higgs contributions),
at $\sqrt s=350$ GeV, for $M_{H^0}=170$ GeV (continuous line),
$M_{H^0}=185$ GeV (dashed line),
$M_{H^0}=210$ GeV (dotted line) and
$M_{H^0}=240$ GeV (chain-dashed line) with $m_t=150$ GeV
and $m_t=175$ GeV, after
the cuts:
$|M_{Z^0}-M_{b\bar b}|<10$ GeV and $|\cos\theta_{b\bar b}|<0.8$.

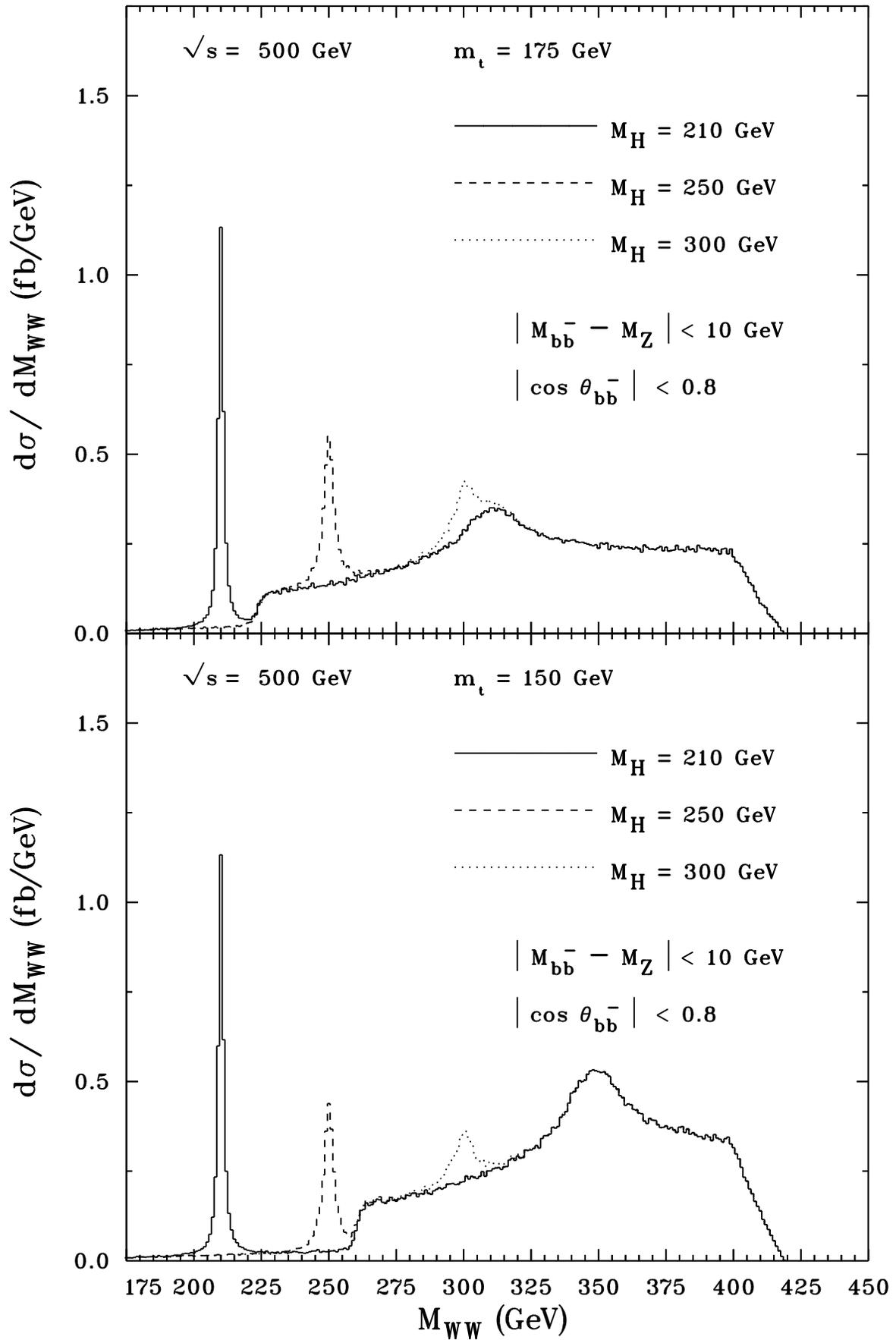
\item[fig.4] The differential distribution
$d\sigma/dM_{W^+W^-}$ for
$e^+e^-\rightarrow b\bar b W^+W^-$ (full matrix element
with all Higgs contributions),
at $\sqrt s=500$ GeV, for $M_{H^0}=210$ GeV (continuous line),
$M_{H^0}=250$ GeV (dashed line)
and $M_{H^0}=300$ GeV (dotted line), with $m_t=150$ GeV
and $m_t=175$ GeV, after
the cuts:
$|M_{Z^0}-M_{b\bar b}|<10$ GeV and $|\cos\theta_{b\bar b}|<0.8$.

\end{description}
\vfill
%%%%%%%%%%%%%%%%%%%%%%%%%%%%%%%%%%%%%%
%\end{document}
%%%%%%%%%%%%%%%%%%%%%%%%%%%%%%%%%%%%%%
\newpage
\pagestyle{empty}

\vspace*{-1.5cm}

\begin{table}%[p]%[htbp]
\begin{center}
\begin{tabular}{|c|c|c|c|}
\hline
\rule[-0.5cm]{0cm}{1.1cm}
$M_{H^0}~{\rm{(GeV)}}$ &\makebox[2.2cm]{Signal} & Background
& $S/\sqrt{B}$  \\ \hline\hline
\multicolumn{4}{|c|}
{\rule[-0.5cm]{0cm}{1.1cm}
 $~~\sqrt s=350~{\rm{GeV}}~~m_t=150~{\rm{GeV}}$ }
\\ \hline
\rule[-0.6cm]{0cm}{1.3cm}
$185$ & $40$ & $132$ & $3.5$    \\ %\hline
\rule[-0.6cm]{0cm}{.9cm}
$210$ & $23$ & $80$ & $2.6$     \\  %\hline
\rule[-0.6cm]{0cm}{.9cm}
$240$ & $10.5$ & $71$ & $1.3$     \\  \hline\hline
\multicolumn{4}{|c|}
{\rule[-0.5cm]{0cm}{1.1cm}
$~~\sqrt s=500~{\rm{GeV}}~~m_t=150~{\rm{GeV}}$ }
 \\ \hline
\rule[-0.6cm]{0cm}{.9cm}
$250$ & $11.5$ & $1.5$ & $9.1$    \\  %\hline
\rule[-0.6cm]{0cm}{.9cm}
$300$ & $5.5$ & $13.5$ & $1.5$     \\ \hline\hline
\multicolumn{4}{|c|}
{\rule[-0.5cm]{0cm}{1.1cm}
$~~\sqrt s=500~{\rm{GeV}}~~m_t=175~{\rm{GeV}}$ }
 \\ \hline
\rule[-0.6cm]{0cm}{.9cm}
$250$ & $11.5$ & $8$ & $4.0$    \\  %\hline
\rule[-0.6cm]{0cm}{.9cm}
$300$ & $5.5$ & $17$ & $1.4$     \\ \hline\hline
\multicolumn{4}{|c|}
{\rule[-0.5cm]{0cm}{1.1cm}
${\cal L} = 10~fb^{-1}~~~~\epsilon_b=1/3$ }
 \\ \hline
\multicolumn{4}{c}
{\rule{0cm}{.9cm}
{\Large Table I}}  \\
\multicolumn{4}{c}
{\rule{0cm}{.9cm}}

\end{tabular}
\end{center}
\end{table}

\newpage
\pagestyle{empty}

\vspace*{-1.5cm}

\begin{table}%[p]%[htbp]
\begin{center}
\begin{tabular}{|c|c|c|c|}     \hline
\rule[-0.5cm]{0cm}{1.1cm}
$M_{H^0}~{\rm{(GeV)}}$ & \omit $~$ & \omit
$\sigma~{\rm (fb)}~~~~~~~~$
& $~$  \\ \hline
\rule[-0.6cm]{0cm}{1.1cm}
&$Z^0H^0\rightarrow b\bar b W^+W^-$
 & $Z^0H^0\rightarrow b\bar b W^+W^-$
& $b\bar bW^+W^-$  \\[-.5 cm]
  & no cut & cut & \\ \hline
\rule[-0.6cm]{0cm}{1.3cm}
$170$ & $13.24$ & $10.61$ & $1.11$    \\ %\hline
\rule[-0.6cm]{0cm}{.9cm}
$180$ & $12.07$ & $8.74$ & $1.35$     \\  %\hline
\rule[-0.6cm]{0cm}{.9cm}
$190$ & $8.46$ & $5.88$ & $1.44$     \\  %\hline
\rule[-0.6cm]{0cm}{.9cm}
$200$ & $7.05$ & $4.99$ & $1.48$    \\  %\hline
\rule[-0.6cm]{0cm}{.9cm}
$210$ & $5.94$ & $4.31$ & $1.42$     \\ \hline
\rule[-0.5cm]{0cm}{1.1cm}
$\;\;\;\;\;$ & \omit $~~\sqrt s=350~{\rm{GeV}}~~$ & \omit $\;\;\;\;\;$
& $m_t=150~{\rm{GeV}}$  \\ \hline
\multicolumn{4}{c}
{\rule{0cm}{.9cm}
{\Large Table II}}  \\
\multicolumn{4}{c}
{\rule{0cm}{.9cm}}

\end{tabular}

\begin{tabular}{|c|c|c|c|}     \hline
\rule[-0.5cm]{0cm}{1.1cm}
$M_{H^0}~{\rm{(GeV)}}$ & \omit $~$ & \omit
$\sigma~{\rm (fb)}~~~~~~~~$
& $~$  \\ \hline
\rule[-0.6cm]{0cm}{1.1cm}
& $Z^0H^0\rightarrow b\bar b W^+W^-$  & $Z^0H^0\rightarrow b\bar b W^+W^-$
& $b\bar bW^+W^-$  \\[-.5 cm]
  & no cut & cut & \\ \hline
\rule[-0.6cm]{0cm}{1.3cm}
$180$ & $6.89$ & $6.23$  & $0.42$    \\%\hline
\rule[-0.6cm]{0cm}{.9cm}
$210$ & $4.35$ & $3.79$  & $0.54$     \\  %\hline
\rule[-0.6cm]{0cm}{.9cm}
$240$ & $3.55$ & $2.94$  & $0.73$     \\  %\hline
\rule[-0.6cm]{0cm}{.9cm}
$270$ & $2.80$ & $2.15$  & $0.84$    \\  %\hline
\rule[-0.6cm]{0cm}{.9cm}
$300$ & $2.10$ & $1.46$  & $0.93$     \\  \hline
\rule[-0.5cm]{0cm}{1.1cm}
$\;\;\;\;\;$ & \omit $~~\sqrt s=500~{\rm{GeV}}~~$ & \omit $\;\;\;\;\;$
& $m_t=150~{\rm{GeV}}$  \\ \hline
\multicolumn{4}{c}
{\rule{0cm}{.9cm}
{\Large Table III}}
\end{tabular}

\end{center}
\end{table}

%\end{document}

\newpage

\
\vskip 2.0cm

\begin{picture}(10000,8000)
\THICKLINES
\bigphotons
\drawline\photon[\W\REG](10000,8000)[6]
\drawline\fermion[\NW\REG](\photonbackx,\photonbacky)[5000]
\drawarrow[\SE\ATBASE](\pmidx,\pmidy)
\drawline\fermion[\SW\REG](\photonbackx,\photonbacky)[5000]
\drawarrow[\SW\ATBASE](\pmidx,\pmidy)
\drawline\fermion[\NE\REG](\photonfrontx,\photonfronty)[5000]
\drawarrow[\NE\ATBASE](\pmidx,\pmidy)
\drawline\photon[\E\REG](12500,10500)[3]
\drawline\photon[\E\REG](10500,8500)[5]
\drawline\fermion[\SE\REG](10000,8000)[5000]
\drawarrow[\NW\ATBASE](\pmidx,\pmidy)
\put(-500,12000){$e^-$}
\put(-500,3000){$e^+$}
\put(14000,12000){$b$}
\put(14000,3000){$\bar b$}
\put(16000,10000){$W^+$}
\put(16000,8000){$W^-$}
\put(6500,2000){$(1)$}
\drawline\photon[\W\REG](32000,8000)[6]
\drawline\fermion[\NW\REG](\photonbackx,\photonbacky)[5000]
\drawarrow[\SE\ATBASE](\pmidx,\pmidy)
\drawline\fermion[\SW\REG](\photonbackx,\photonbacky)[5000]
\drawarrow[\SW\ATBASE](\pmidx,\pmidy)
\drawline\fermion[\NE\REG](\photonfrontx,\photonfronty)[5000]
\drawarrow[\NE\ATBASE](\pmidx,\pmidy)
\drawline\photon[\E\REG](\pmidx,\pmidy)[4]
\drawline\fermion[\SE\REG](32000,8000)[5000]
\drawarrow[\NW\ATBASE](\pmidx,\pmidy)
\drawline\photon[\E\REG](\pmidx,\pmidy)[4]
\put(21500,12000){$e^-$}
\put(21500,3000){$e^+$}
\put(36000,12000){$b$}
\put(36000,3000){$\bar b$}
\put(38250,9250){$W^+$}
\put(38250,5750){$W^-$}
\put(28500,2000){$(2)$}
\end{picture}

\vskip 2.0cm

\begin{picture}(10000,8000)
\THICKLINES
\bigphotons
\drawline\photon[\W\REG](10000,8000)[6]
\drawline\fermion[\NW\REG](\photonbackx,\photonbacky)[5000]
\drawarrow[\SE\ATBASE](\pmidx,\pmidy)
\drawline\fermion[\SW\REG](\photonbackx,\photonbacky)[5000]
\drawarrow[\SW\ATBASE](\pmidx,\pmidy)
\drawline\fermion[\NE\REG](\photonfrontx,\photonfronty)[5000]
\drawarrow[\NE\ATBASE](\pmidx,\pmidy)
\drawline\fermion[\SE\REG](10000,8000)[5000]
\drawarrow[\NW\ATBASE](\pmidx,\pmidy)
\drawline\photon[\E\REG](12500,5500)[3]
\drawline\photon[\E\REG](10500,7500)[5]
\put(-500,12000){$e^-$}
\put(-500,3000){$e^+$}
\put(14000,12000){$b$}
\put(14000,3000){$\bar b$}
\put(16000,5000){$W^-$}
\put(16000,7000){$W^+$}
\put(6500,2000){$(3)$}
\drawline\photon[\W\REG](32000,8000)[6]
\drawline\fermion[\NW\REG](\photonbackx,\photonbacky)[5000]
\drawarrow[\SE\ATBASE](\pmidx,\pmidy)
\drawline\fermion[\SW\REG](\photonbackx,\photonbacky)[5000]
\drawarrow[\SW\ATBASE](\pmidx,\pmidy)
\drawline\photon[\E\REG](23000,11000)[5]
\drawline\photon[\E\REG](25000,9000)[3]
\drawline\fermion[\NE\REG](32000,8000)[5000]
\drawarrow[\NE\ATBASE](\pmidx,\pmidy)
\drawline\fermion[\SE\REG](32000,8000)[5000]
\drawarrow[\NW\ATBASE](\pmidx,\pmidy)
\put(21500,12000){$e^-$}
\put(21500,3000){$e^+$}
\put(36000,12000){$b$}
\put(36000,3000){$\bar b$}
\put(28500,11000){$W^-$}
\put(28500,9000){$W^+$}
\put(28500,2000){$(4)$}
\end{picture}

\vskip 2.0cm

\begin{picture}(10000,8000)
\THICKLINES
\bigphotons
\drawline\photon[\W\REG](10000,8000)[6]
\drawline\fermion[\NW\REG](\photonbackx,\photonbacky)[5000]
\drawarrow[\SE\ATBASE](\pmidx,\pmidy)
\drawline\photon[\E\REG](\pmidx,\pmidy)[4]
\drawline\fermion[\SW\REG](4000,8000)[5000]
\drawarrow[\SW\ATBASE](\pmidx,\pmidy)
\drawline\photon[\E\REG](\pmidx,\pmidy)[4]
\drawline\fermion[\NE\REG](10000,8000)[5000]
\drawarrow[\NE\ATBASE](\pmidx,\pmidy)
\drawline\fermion[\SE\REG](10000,8000)[5000]
\drawarrow[\NW\ATBASE](\pmidx,\pmidy)
\put(-500,12000){$e^-$}
\put(-500,3000){$e^+$}
\put(14000,12000){$b$}
\put(14000,3000){$\bar b$}
\put(6500,9500){$W^-$}
\put(6500,5500){$W^+$}
\put(6500,2000){$(5)$}
\drawline\photon[\W\REG](32000,8000)[6]
\drawline\fermion[\NW\REG](\photonbackx,\photonbacky)[5000]
\drawarrow[\SE\ATBASE](\pmidx,\pmidy)
\drawline\photon[\E\REG](23000,5000)[5]
\drawline\photon[\E\REG](25000,7000)[3]
\drawline\fermion[\SW\REG](26000,8000)[5000]
\drawarrow[\SW\ATBASE](\pmidx,\pmidy)
\drawline\fermion[\NE\REG](32000,8000)[5000]
\drawarrow[\NE\ATBASE](\pmidx,\pmidy)
\drawline\fermion[\SE\REG](32000,8000)[5000]
\drawarrow[\NW\ATBASE](\pmidx,\pmidy)
\put(21500,12000){$e^-$}
\put(21500,3000){$e^+$}
\put(36000,12000){$b$}
\put(36000,3000){$\bar b$}
\put(28500,4500){$W^+$}
\put(28500,6500){$W^-$}
\put(28500,2000){$(6)$}
\end{picture}

\vskip 2.0cm

\begin{picture}(10000,8000)
\THICKLINES
\bigphotons
\drawline\photon[\W\REG](10000,8000)[6]
\drawline\fermion[\NW\REG](\photonbackx,\photonbacky)[5000]
\drawarrow[\SE\ATBASE](\pmidx,\pmidy)
\drawline\photon[\E\REG](\pmidx,\pmidy)[4]
\drawline\fermion[\SW\REG](4000,8000)[5000]
\drawarrow[\SW\ATBASE](\pmidx,\pmidy)
\drawline\fermion[\NE\REG](10000,8000)[5000]
\drawarrow[\NE\ATBASE](\pmidx,\pmidy)
\drawline\photon[\E\REG](\pmidx,\pmidy)[4]
\drawline\fermion[\SE\REG](10000,8000)[5000]
\drawarrow[\NW\ATBASE](\pmidx,\pmidy)
\put(-500,12000){$e^-$}
\put(-500,3000){$e^+$}
\put(14000,12000){$b$}
\put(14000,3000){$\bar b$}
\put(16000,9500){$W^+$}
\put(6500,9500){$W^-$}
\put(6500,2000){$(7)$}
\drawline\photon[\W\REG](32000,8000)[6]
\drawline\fermion[\NW\REG](\photonbackx,\photonbacky)[5000]
\drawarrow[\SE\ATBASE](\pmidx,\pmidy)
\drawline\fermion[\SW\REG](26000,8000)[5000]
\drawarrow[\SW\ATBASE](\pmidx,\pmidy)
\drawline\photon[\E\REG](\pmidx,\pmidy)[4]
\drawline\fermion[\NE\REG](32000,8000)[5000]
\drawarrow[\NE\ATBASE](\pmidx,\pmidy)
\drawline\fermion[\SE\REG](32000,8000)[5000]
\drawarrow[\NW\ATBASE](\pmidx,\pmidy)
\drawline\photon[\E\REG](\pmidx,\pmidy)[4]
\put(21500,12000){$e^-$}
\put(21500,3000){$e^+$}
\put(36000,12000){$b$}
\put(36000,3000){$\bar b$}
\put(38250,5500){$W^-$}
\put(28500,5500){$W^+$}
\put(28500,2000){$(8)$}
\end{picture}

\vskip 1.0cm
\centerline{\bf\Large Fig.~1}

\vfill
\newpage
\pagestyle{empty}
\
\vskip 2.0cm

\begin{picture}(10000,8000)
\THICKLINES
\bigphotons
\drawline\photon[\W\REG](10000,8000)[6]
\drawline\fermion[\NW\REG](\photonbackx,\photonbacky)[5000]
\drawarrow[\SE\ATBASE](\pmidx,\pmidy)
\drawline\fermion[\SW\REG](\photonbackx,\photonbacky)[5000]
\drawarrow[\SW\ATBASE](\pmidx,\pmidy)
\drawline\fermion[\NE\REG](\photonfrontx,\photonfronty)[5000]
\drawarrow[\NE\ATBASE](\pmidx,\pmidy)
\drawvertex\photon[\E 3](\pmidx,\pmidy)[3]
\drawline\fermion[\SE\REG](10000,8000)[5000]
\drawarrow[\NW\ATBASE](\pmidx,\pmidy)
\put(-500,12000){$e^-$}
\put(-500,3000){$e^+$}
\put(14000,12000){$b$}
\put(14000,3000){$\bar b$}
\put(17300,11700){$W^+$}
\put(17300,7050){$W^-$}
\put(6500,2000){$(9)$}
\drawline\photon[\W\REG](32000,8000)[6]
\drawline\fermion[\NW\REG](\photonbackx,\photonbacky)[5000]
\drawarrow[\SE\ATBASE](\pmidx,\pmidy)
\drawline\fermion[\SW\REG](\photonbackx,\photonbacky)[5000]
\drawarrow[\SW\ATBASE](\pmidx,\pmidy)
\drawline\fermion[\NE\REG](\photonfrontx,\photonfronty)[5000]
\drawarrow[\NE\ATBASE](\pmidx,\pmidy)
\drawline\fermion[\SE\REG](32000,8000)[5000]
\drawarrow[\NW\ATBASE](\pmidx,\pmidy)
\drawvertex\photon[\E 3](\pmidx,\pmidy)[3]
\put(21500,12000){$e^-$}
\put(21500,3000){$e^+$}
\put(36000,12000){$b$}
\put(36000,3000){$\bar b$}
\put(39300,8200){$W^+$}
\put(39300,3550){$W^-$}
\put(28000,2000){$(10)$}
\end{picture}

\vskip 2.0cm

\begin{picture}(10000,8000)
\THICKLINES
\bigphotons
\drawline\photon[\W\REG](11000,0)[4]
\drawline\fermion[\N\REG](\photonbackx,\photonbacky)[8000]
\drawarrow[\S\ATBASE](\pmidx,\pmidy)
\drawline\fermion[\NW\REG](\fermionbackx,\fermionbacky)[5000]
\drawarrow[\SE\ATBASE](\pmidx,\pmidy)
\drawline\fermion[\SW\REG](7000,0)[5000]
\drawarrow[\SW\ATBASE](\pmidx,\pmidy)
\drawline\fermion[\NE\REG](11000,0)[4000]
\drawarrow[\NE\ATBASE](\pmidx,\pmidy)
\drawline\fermion[\SE\REG](\fermionfrontx,\fermionfronty)[4000]
\drawarrow[\NW\ATBASE](\pmidx,\pmidy)
\drawvertex\photon[\E 3](7000,8000)[4]
\put(2500,12000){$e^-$}
\put(2500,-5000){$e^+$}
\put(14150,10500){$W^+$}
\put(14150,4750){$W^-$}
\put(14300,2750){$b$}
\put(14300,-3750){$\bar b$}
\put(8000,-7000){$(11)$}
\drawline\photon[\W\REG](33000,8000)[4]
\drawline\fermion[\NW\REG](\photonbackx,\photonbacky)[5000]
\drawarrow[\SE\ATBASE](\pmidx,\pmidy)
\drawline\fermion[\S\REG](\photonbackx,\photonbacky)[8000]
\drawarrow[\S\ATBASE](\pmidx,\pmidy)
\drawline\fermion[\SW\REG](\fermionbackx,\fermionbacky)[5000]
\drawarrow[\SW\ATBASE](\pmidx,\pmidy)
\drawline\fermion[\NE\REG](\photonfrontx,\photonfronty)[4000]
\drawarrow[\NE\ATBASE](\pmidx,\pmidy)
\drawline\fermion[\SE\REG](33000,8000)[4000]
\drawarrow[\NW\ATBASE](\pmidx,\pmidy)
\drawvertex\photon[\E 3](29000,0)[4]
\put(24500,12000){$e^-$}
\put(24500,-5000){$e^+$}
\put(36300,10750){$b$}
\put(36300,4250){$\bar b$}
\put(36150,2500){$W^+$}
\put(36150,-3250){$W^-$}
\put(30000,-7000){$(12)$}
\end{picture}

\vskip 3.0cm
\centerline{\bf\Large Fig.~1 (Continued)}

\vfill
\newpage
\pagestyle{empty}
\
\vskip 2.0cm

\begin{picture}(10000,8000)
\THICKLINES
\bigphotons
\drawline\photon[\W\REG](10000,8000)[6]
\drawline\fermion[\NW\REG](\photonbackx,\photonbacky)[5000]
\drawarrow[\SE\ATBASE](\pmidx,\pmidy)
\drawline\fermion[\SW\REG](\photonbackx,\photonbacky)[5000]
\drawarrow[\SW\ATBASE](\pmidx,\pmidy)
\drawline\photon[\NE\REG](\photonfrontx,\photonfronty)[6]
\drawline\photon[\E\REG](12300,10000)[3]
\drawline\fermion[\NE\REG](\photonbackx,\photonbacky)[3000]
\drawarrow[\NE\ATBASE](\pmidx,\pmidy)
\drawline\fermion[\SE\REG](\photonbackx,\photonbacky)[3000]
\drawarrow[\NW\ATBASE](\pmidx,\pmidy)
\drawline\photon[\SE\REG](10000,8000)[6]
\put(-500,12000){$e^-$}
\put(-500,3000){$e^+$}
\put(14000,12000){$W^+$}
\put(14000,3000){$W^-$}
\put(17650,12000){$b$}
\put(17650,6750){$\bar b$}
\put(6000,2000){$(13)$}
\drawline\photon[\W\REG](32000,8000)[6]
\drawline\fermion[\NW\REG](\photonbackx,\photonbacky)[5000]
\drawarrow[\SE\ATBASE](\pmidx,\pmidy)
\drawline\fermion[\SW\REG](\photonbackx,\photonbacky)[5000]
\drawarrow[\SW\ATBASE](\pmidx,\pmidy)
\drawline\photon[\NE\REG](\photonfrontx,\photonfronty)[6]
\drawline\photon[\SE\REG](32000,8000)[6]
\drawline\photon[\E\REG](\pmidx,\pmidy)[3]
\drawline\fermion[\NE\REG](\photonbackx,\photonbacky)[3000]
\drawarrow[\NE\ATBASE](\pmidx,\pmidy)
\drawline\fermion[\SE\REG](\photonbackx,\photonbacky)[3000]
\drawarrow[\NW\ATBASE](\pmidx,\pmidy)
\put(21500,12000){$e^-$}
\put(21500,3000){$e^+$}
\put(36000,12000){$W^+$}
\put(36000,3000){$W^-$}
\put(39300,8400){$b$}
\put(39300,2900){$\bar b$}
\put(28000,2000){$(14)$}
\end{picture}

\vskip 2.0cm

\begin{picture}(10000,8000)
\THICKLINES
\bigphotons
\drawline\photon[\W\REG](21000,8000)[5]
\drawline\fermion[\NW\REG](\photonbackx,\photonbacky)[5000]
\drawarrow[\SE\ATBASE](\pmidx,\pmidy)
\drawline\fermion[\SW\REG](\photonbackx,\photonbacky)[5000]
\drawarrow[\SW\ATBASE](\pmidx,\pmidy)
\drawline\photon[\NE\REG](\photonfrontx,\photonfronty)[6]
\drawline\photon[\E\REG](\photonfrontx,\photonfronty)[5]
\drawline\fermion[\NE\REG](\photonbackx,\photonbacky)[3000]
\drawarrow[\NE\ATBASE](\pmidx,\pmidy)
\drawline\fermion[\SE\REG](\photonbackx,\photonbacky)[3000]
\drawarrow[\NW\ATBASE](\pmidx,\pmidy)
\drawline\photon[\SE\REG](21000,8000)[6]
\put(11500,12000){$e^-$}
\put(11500,3000){$e^+$}
\put(25000,12000){$W^+$}
\put(25000,3000){$W^-$}
\put(28400,10250){$b$}
\put(28400,4750){$\bar b$}
\put(20000,2000){$(15)$}
\end{picture}

%\vskip 1.0cm
%\centerline{\bf\Large Fig.~1 (Continued)}

%\vfill
%\newpage
%\pagestyle{empty}
%\
\vskip 2.0cm

\begin{picture}(10000,8000)
\THICKLINES
\bigphotons
\drawline\photon[\W\REG](9000,8000)[5]
\drawline\fermion[\NW\REG](\photonbackx,\photonbacky)[5000]
\drawarrow[\SE\ATBASE](\pmidx,\pmidy)
\drawline\fermion[\SW\REG](\photonbackx,\photonbacky)[5000]
\drawarrow[\SW\ATBASE](\pmidx,\pmidy)
\drawline\photon[\NE\REG](\photonfrontx,\photonfronty)[6]
\drawline\photon[\SE\REG](9000,8000)[4]
\drawline\fermion[\NE\REG](\photonbackx,\photonbacky)[3000]
\drawarrow[\NE\ATBASE](\pmidx,\pmidy)
\drawline\fermion[\SE\REG](\photonbackx,\photonbacky)[3000]
\drawarrow[\NW\ATBASE](\pmidx,\pmidy)
\drawline\photon[\E\REG](\pmidx,\pmidy)[3]
\put(-500,12000){$e^-$}
\put(-500,3000){$e^+$}
\put(13000,12000){$W^+$}
\put(15900,4200){$W^-$}
\put(13750,8000){$b$}
\put(13750,2250){$\bar b$}
\put(6000,2000){$(16)$}
\drawline\photon[\W\REG](31000,8000)[5]
\drawline\fermion[\NW\REG](\photonbackx,\photonbacky)[5000]
\drawarrow[\SE\ATBASE](\pmidx,\pmidy)
\drawline\fermion[\SW\REG](\photonbackx,\photonbacky)[5000]
\drawarrow[\SW\ATBASE](\pmidx,\pmidy)
\drawline\photon[\NE\REG](\photonfrontx,\photonfronty)[4]
\drawline\fermion[\NE\REG](\photonbackx,\photonbacky)[3000]
\drawarrow[\NE\ATBASE](\pmidx,\pmidy)
\drawline\photon[\E\REG](\pmidx,\pmidy)[3]
\drawline\fermion[\SE\REG](\fermionfrontx,\fermionfronty)[3000]
\drawarrow[\NW\ATBASE](\pmidx,\pmidy)
\drawline\photon[\SE\REG](31000,8000)[6]
\put(21500,12000){$e^-$}
\put(21500,3000){$e^+$}
\put(37800,11250){$W^+$}
\put(35000,3000){$W^-$}
\put(35750,12750){$b$}
\put(35750,7000){$\bar b$}
\put(28000,2000){$(17)$}
\end{picture}

\vskip 2.0cm

\begin{picture}(10000,8000)
\THICKLINES
\bigphotons
\drawline\photon[\W\REG](9000,8000)[5]
\drawline\fermion[\NW\REG](\photonbackx,\photonbacky)[5000]
\drawarrow[\SE\ATBASE](\pmidx,\pmidy)
\drawline\photon[\E\REG](\pmidx,\pmidy)[3]
\drawline\fermion[\SW\REG](\fermionfrontx,\fermionfronty)[5000]
\drawarrow[\SW\ATBASE](\pmidx,\pmidy)
\drawline\photon[\NE\REG](9000,8000)[6]
\drawline\photon[\SE\REG](9000,8000)[4]
\drawline\fermion[\NE\REG](\photonbackx,\photonbacky)[3000]
\drawarrow[\NE\ATBASE](\pmidx,\pmidy)
\drawline\fermion[\SE\REG](\photonbackx,\photonbacky)[3000]
\drawarrow[\NW\ATBASE](\pmidx,\pmidy)
\put(-500,12000){$e^-$}
\put(-500,3000){$e^+$}
\put(13000,12000){$W^+$}
\put(5500,9500){$W^-$}
\put(13750,8000){$b$}
\put(13750,2250){$\bar b$}
\put(6000,2000){$(18)$}
\drawline\photon[\W\REG](31000,8000)[5]
\drawline\fermion[\NW\REG](\photonbackx,\photonbacky)[5000]
\drawarrow[\SE\ATBASE](\pmidx,\pmidy)
\drawline\fermion[\SW\REG](\fermionfrontx,\fermionfronty)[5000]
\drawarrow[\SW\ATBASE](\pmidx,\pmidy)
\drawline\photon[\E\REG](\pmidx,\pmidy)[3]
\drawline\photon[\SE\REG](31000,8000)[6]
\drawline\photon[\NE\REG](31000,8000)[4]
\drawline\fermion[\NE\REG](\photonbackx,\photonbacky)[3000]
\drawarrow[\NE\ATBASE](\pmidx,\pmidy)
\drawline\fermion[\SE\REG](\photonbackx,\photonbacky)[3000]
\drawarrow[\NW\ATBASE](\pmidx,\pmidy)
\put(21500,12000){$e^-$}
\put(21500,3000){$e^+$}
\put(35000,3000){$W^-$}
\put(27500,5750){$W^+$}
\put(35750,12500){$b$}
\put(35750,7000){$\bar b$}
\put(28000,2000){$(19)$}
\end{picture}

\vskip 1.0cm
\centerline{\bf\Large Fig.~1 (Continued)}

\vfill
\newpage
\pagestyle{empty}
\
\vskip 2.0cm

\begin{picture}(10000,8000)
\THICKLINES
\bigphotons
\drawline\photon[\W\REG](10000,8000)[6]
\drawline\fermion[\NW\REG](\photonbackx,\photonbacky)[5000]
\drawarrow[\SE\ATBASE](\pmidx,\pmidy)
\drawline\fermion[\SW\REG](\photonbackx,\photonbacky)[5000]
\drawarrow[\SW\ATBASE](\pmidx,\pmidy)
\drawline\fermion[\NE\REG](\photonfrontx,\photonfronty)[5000]
\drawarrow[\NE\ATBASE](\pmidx,\pmidy)
\seglength=1416  \gaplength=300  % Changes the \scalar defaults.
\drawline\scalar[\E\REG](\pmidx,\pmidy)[2]
\drawline\photon[\NE\REG](\scalarbackx,\scalarbacky)[3]
\drawline\photon[\SE\REG](\scalarbackx,\scalarbacky)[3]
\drawline\fermion[\SE\REG](10000,8000)[5000]
\drawarrow[\NW\ATBASE](\pmidx,\pmidy)
\put(-500,12000){$e^-$}
\put(-500,3000){$e^+$}
\put(14000,12000){$b$}
\put(14000,3000){$\bar b$}
\put(17250,11250){$W^+$}
\put(17250,7500){$W^-$}
\put(6000,2000){$(20)$}
\drawline\photon[\W\REG](32000,8000)[6]
\drawline\fermion[\NW\REG](\photonbackx,\photonbacky)[5000]
\drawarrow[\SE\ATBASE](\pmidx,\pmidy)
\drawline\fermion[\SW\REG](\photonbackx,\photonbacky)[5000]
\drawarrow[\SW\ATBASE](\pmidx,\pmidy)
\drawline\fermion[\NE\REG](\photonfrontx,\photonfronty)[5000]
\drawarrow[\NE\ATBASE](\pmidx,\pmidy)
\drawline\fermion[\SE\REG](32000,8000)[5000]
\drawarrow[\NW\ATBASE](\pmidx,\pmidy)
\seglength=1416  \gaplength=300  % Changes the \scalar defaults.
\drawline\scalar[\E\REG](\pmidx,\pmidy)[2]
\drawline\photon[\NE\REG](\scalarbackx,\scalarbacky)[3]
\drawline\photon[\SE\REG](\scalarbackx,\scalarbacky)[3]
\put(21500,12000){$e^-$}
\put(21500,3000){$e^+$}
\put(36000,12000){$b$}
\put(36000,3000){$\bar b$}
\put(39250,7750){$W^+$}
\put(39250,4000){$W^-$}
\put(28000,2000){$(21)$}
\end{picture}

\vskip 2.0cm

\begin{picture}(10000,8000)
\THICKLINES
\bigphotons
\drawline\photon[\W\REG](10000,8000)[6]
\drawline\fermion[\NW\REG](\photonbackx,\photonbacky)[5000]
\drawarrow[\SE\ATBASE](\pmidx,\pmidy)
\drawline\fermion[\SW\REG](\photonbackx,\photonbacky)[5000]
\drawarrow[\SW\ATBASE](\pmidx,\pmidy)
\drawline\photon[\NE\REG](\photonfrontx,\photonfronty)[6]
\seglength=1416  \gaplength=300  % Changes the \scalar defaults.
\drawline\scalar[\E\REG](\pmidx,\pmidy)[2]
\drawline\fermion[\NE\REG](\scalarbackx,\scalarbacky)[3000]
\drawarrow[\NE\ATBASE](\pmidx,\pmidy)
\drawline\fermion[\SE\REG](\scalarbackx,\scalarbacky)[3000]
\drawarrow[\NW\ATBASE](\pmidx,\pmidy)
\drawline\photon[\SE\REG](10000,8000)[6]
\put(-500,12000){$e^-$}
\put(-500,3000){$e^+$}
\put(14000,12000){$W^+$}
\put(14000,3000){$W^-$}
\put(17500,12250){$b$}
\put(17500,6750){$\bar b$}
\put(6000,2000){$(22)$}
\drawline\photon[\W\REG](32000,8000)[6]
\drawline\fermion[\NW\REG](\photonbackx,\photonbacky)[5000]
\drawarrow[\SE\ATBASE](\pmidx,\pmidy)
\drawline\fermion[\SW\REG](\photonbackx,\photonbacky)[5000]
\drawarrow[\SW\ATBASE](\pmidx,\pmidy)
\drawline\photon[\NE\REG](\photonfrontx,\photonfronty)[6]
\drawline\photon[\SE\REG](32000,8000)[6]
\seglength=1416  \gaplength=300  % Changes the \scalar defaults.
\drawline\scalar[\E\REG](\pmidx,\pmidy)[2]
\drawline\fermion[\NE\REG](\scalarbackx,\scalarbacky)[3000]
\drawarrow[\NE\ATBASE](\pmidx,\pmidy)
\drawline\fermion[\SE\REG](\scalarbackx,\scalarbacky)[3000]
\drawarrow[\NW\ATBASE](\pmidx,\pmidy)
\put(21500,12000){$e^-$}
\put(21500,3000){$e^+$}
\put(36000,12000){$W^+$}
\put(36000,3000){$W^-$}
\put(39500,8500){$b$}
\put(39500,3000){$\bar b$}
\put(28000,2000){$(23)$}
\end{picture}

\vskip 2.0cm

\begin{picture}(10000,8000)
\THICKLINES
\bigphotons
\drawline\photon[\W\REG](9000,8000)[5]
\drawline\fermion[\NW\REG](\photonbackx,\photonbacky)[5000]
\drawarrow[\SE\ATBASE](\pmidx,\pmidy)
\drawline\photon[\E\REG](\pmidx,\pmidy)[3]
\drawline\fermion[\SW\REG](\fermionfrontx,\fermionfronty)[5000]
\drawarrow[\SW\ATBASE](\pmidx,\pmidy)
\drawline\photon[\NE\REG](9000,8000)[6]
\seglength=1416  \gaplength=300  % Changes the \scalar defaults.
\drawline\scalar[\SE\REG](\photonfrontx,\photonfronty)[2]
\drawline\fermion[\NE\REG](\scalarbackx,\scalarbacky)[3000]
\drawarrow[\NE\ATBASE](\pmidx,\pmidy)
\drawline\fermion[\SE\REG](\scalarbackx,\scalarbacky)[3000]
\drawarrow[\NW\ATBASE](\pmidx,\pmidy)
\put(-500,12000){$e^-$}
\put(-500,3000){$e^+$}
\put(13000,12000){$W^+$}
\put(5650,9500){$W^-$}
\put(13850,8000){$b$}
\put(13850,2750){$\bar b$}
\put(6000,2000){$(24)$}
\drawline\photon[\W\REG](31000,8000)[5]
\drawline\fermion[\NW\REG](\photonbackx,\photonbacky)[5000]
\drawarrow[\SE\ATBASE](\pmidx,\pmidy)
\drawline\fermion[\SW\REG](\fermionfrontx,\fermionfronty)[5000]
\drawarrow[\SW\ATBASE](\pmidx,\pmidy)
\drawline\photon[\E\REG](\pmidx,\pmidy)[3]
\drawline\photon[\SE\REG](31000,8000)[6]
\seglength=1416  \gaplength=300  % Changes the \scalar defaults.
\drawline\scalar[\NE\REG](\photonfrontx,\photonfronty)[2]
\drawline\fermion[\NE\REG](\scalarbackx,\scalarbacky)[3000]
\drawarrow[\NE\ATBASE](\pmidx,\pmidy)
\drawline\fermion[\SE\REG](\scalarbackx,\scalarbacky)[3000]
\drawarrow[\NW\ATBASE](\pmidx,\pmidy)
\put(21500,12000){$e^-$}
\put(21500,3000){$e^+$}
\put(35000,3000){$W^-$}
\put(27650,5900){$W^+$}
\put(35750,12250){$b$}
\put(35750,6750){$\bar b$}
\put(28000,2000){$(25)$}
\end{picture}

\vskip 2.0cm

\begin{picture}(10000,8000)
\THICKLINES
\bigphotons
\drawline\photon[\W\REG](9000,8000)[5]
\drawline\fermion[\NW\REG](\photonbackx,\photonbacky)[5000]
\drawarrow[\SE\ATBASE](\pmidx,\pmidy)
\drawline\fermion[\SW\REG](\photonbackx,\photonbacky)[5000]
\drawarrow[\SW\ATBASE](\pmidx,\pmidy)
\seglength=1416  \gaplength=300  % Changes the \scalar defaults.
\drawline\scalar[\NE\REG](\photonfrontx,\photonfronty)[3]
\drawline\photon[\NE\REG](\scalarbackx,\scalarbacky)[4]
\drawline\photon[\SE\REG](\scalarbackx,\scalarbacky)[4]
\drawline\photon[\SE\REG](9000,8000)[4]
\drawline\fermion[\NE\REG](\photonbackx,\photonbacky)[3000]
\drawarrow[\NE\ATBASE](\pmidx,\pmidy)
\drawline\fermion[\SE\REG](\photonbackx,\photonbacky)[3000]
\drawarrow[\NW\ATBASE](\pmidx,\pmidy)
\put(-500,12000){$e^-$}
\put(-500,3000){$e^+$}
\put(15250,13250){$W^+$}
\put(15250,8500){$W^-$}
\put(13750,8000){$b$}
\put(13750,2250){$\bar b$}
\put(6000,2000){$(26)$}
\drawline\photon[\W\REG](31000,8000)[5]
\drawline\fermion[\NW\REG](\photonbackx,\photonbacky)[5000]
\drawarrow[\SE\ATBASE](\pmidx,\pmidy)
\drawline\fermion[\SW\REG](\photonbackx,\photonbacky)[5000]
\drawarrow[\SW\ATBASE](\pmidx,\pmidy)
\seglength=1416  \gaplength=300  % Changes the \scalar defaults.
\drawline\scalar[\NE\REG](\photonfrontx,\photonfronty)[3]
\drawline\fermion[\NE\REG](\scalarbackx,\scalarbacky)[3000]
\drawarrow[\NE\ATBASE](\pmidx,\pmidy)
\drawline\fermion[\SE\REG](\scalarbackx,\scalarbacky)[3000]
\drawarrow[\NW\ATBASE](\pmidx,\pmidy)
\drawline\photon[\SE\REG](31000,8000)[4]
\drawline\photon[\NE\REG](\photonbackx,\photonbacky)[4]
\drawline\photon[\SE\REG](\photonfrontx,\photonfronty)[4]
\put(21500,12000){$e^-$}
\put(21500,3000){$e^+$}
\put(37000,13500){$b$}
\put(37000,8750){$\bar b$}
\put(36250,7350){$W^+$}
\put(36250,2250){$W^-$}
\put(28000,2000){$(27)$}
\end{picture}

\vskip 1.0cm
\centerline{\bf\Large Fig.~1 (Continued)}

\vfill
\end{document}